\documentstyle[12pt,aaspp4]{article}
\begin{document}
\title{Rotation curves of UMa galaxies in the context of modified 
Newtonian dynamics}
\author{R.H. Sanders and M.A.W. Verheijen}
\affil{Kapteyn Astronomical Institute, Groningen, The Netherlands}

\begin{abstract}  
This is the third in a series of papers in which spiral galaxy rotation
curves are considered in the context of Milgrom's modified dynamics
(MOND).  The present sample of 30 objects is drawn from a complete
sample of galaxies in the Ursa Major cluster with photometric data by
Tully et al.  (1996) and 21 cm line data by Verheijen (1997).  The
galaxies are roughly all at the same distance (15 to 16 Mpc).  The radio
observations are made with the Westerbork Synthesis Array which means
that the linear resolution of all rotation curves is comparable.  The
greatest advantage of this sample is the existance of K$^{'}$-band
surface photometry for all galaxies; the near-infrared emission, being
relatively free of the effects of dust absorption and less sensitive to
recent star formation, is a more precise tracer of the mean radial
distribution of the dominant stellar population.  The predicted rotation
curves are calculated from the K$^{'}$-band surface photometry and the
observed distribution of neutral hydrogen using the simple MOND
prescription where the one adjustable parameter is the mass of the
stellar disk or the implied mass-to-light ratio.  The predicted rotation
curves generally agree with the observed curves and the mean M/L in the
near-infrared is about 0.9 with a small dispersion.  The fitted M/L in
the B-band is correlated with B-V color in the sense predicted by
population synthesis models.  Including earlier work, about 80 galaxy
rotation curves are now well-reproduced from the observed distribution
of detectable matter using the MOND formula to calculate the
gravitational acceleration; this lends considerable observational
support to Milgrom's unconventional hypothesis. 
\end{abstract}

\section{Introduction}

The rotation curves of spiral galaxies, as observed in the 21 cm line of
neutral hydrogen, provide, in many cases, an accurate determination of
the radial force distribution in regions of very low gravitational
acceleration ($<10^{-8}$ cm/s$^2$).  Therefore, these data are ideal for
testing alternatives to dark matter such as Milgrom's proposed modified
Newtonian dynamics (MOND) which posits that the discrepancy between the
true gravitational force and the Newtonian force appears at low
accelerations.  As a theory of gravity or inertia, MOND predicts the
precise form and amplitude of a rotation curve from the observed radial
distribution of detectable matter (stars and gas) in a spiral galaxy,
often with only one adjustable parameter which is the mass-to-light
ratio of the stellar disk (Milgrom 1983a, b). 

In two previous papers (Begeman et al.  1990 Paper 1, Sanders 1996 Paper
2), a sample of 33 spiral galaxies with published rotation curves has
been considered in the context of MOND.  The 11 galaxies considered in
Paper 1 were highly selected to meet certain strict criteria: the
galaxies were rich in neutral hydrogen; the distribution and velocity
field of the neutral gas was smooth and symmetric; the observed 21 cm
line rotation curves extended far beyond the visible disk; the galaxies
were closer than 1000 km/s to achieve high linear resolution and a large
number of independent points along the rotation curve.  In Paper 2 these
criteria were relaxed to include an additional 22 galaxies with
published 21 cm line rotation curves (as of 1996), but selection was
still essentially based upon HI richness and extent.  For the combined
samples of Papers 1 and 2 it was demonstrated that, in most cases, the
observed rotation curve was predicted in detail from the observed light
and gas distributions using the simple MOND prescription.  Moreover, the
range of fitted values of M/L was astrophysically plausible and
consistent with population synthesis models. 

In the present paper, this work is extended to a new complete and
homogeneous sample of spiral galaxies in the Ursa Major cluster.  The
sample has been previously considered in terms of the distribution by
surface brightness (Tully \& Verheijen 1997), but until now, the
dynamics of these galaxies has not been discussed in the published
literature.  The important aspect of this sample is that it is optically
selected (Tully et al.  1996).  All galaxies in the UMa cluster brighter
than a specified limiting magnitude are considered; although, the final
sample is weighted against gas-poor systems because the rotation curves
are determined from 21 cm line observations made at the Westerbork Radio
Synthesis Telescope (Verheijen 1997).  All sample galaxies have been
imaged in the K$^{'}$-band (2.2 $\mu$m) by Tully et al.  (1996). 
Therefore, this sample is distinguished from the previous samples in its
primary selection criterion and by its homogeneity: rotation curves of
consistent and sufficient linear resolution are combined with
near-infrared surface photometry in a data set which is well-suited to
the purpose of this work. 
   
The previous combined sample of 33 field galaxies (Paper 2) was very
inhomogeneous with the 21 cm line observations being carried out at
either Westerbork or the VLA with differing angular resolution and
sensitivity.  Considering the range of distance covered by the sample
(from 0.8 to 80 Mpc) the actual linear resolution of the neutral
hydrogen observations varied between about 50 pc and 5 kpc.  The 21 cm
line data were taken by different observers using different methods of
analysis.  The photometry was quite inhomogeneous; in most cases this
was CCD photometry but for several objects only older photographic
photometry was available.  The photometry was typically, but not always,
in the B-band which is less reliable than redder bands as a tracer of
the true radial distribution of the dominant stellar population, subject
as it is to recent star-formation and the effects of differential dust
obscuration.  Moreover, nine of these galaxies contained bulges or
central light concentrations which, at least in the B-band, yielded yet
another fitting parameter-- the M/L of the bulge component. 

These problems are minimized in the UMa sample.  All galaxies are
roughly at the same distance (taken to be 15.5 Mpc) which means that the
linear resolution of the 21 cm line observations is similar in all cases
(typically 0.75 kpc).  The relatively low scatter in the observed
Tully-Fisher relation implies a rather small dispersion in distance. 
The observations are reduced in the same way and the same method is
applied in deriving the rotation curves from the line data (Verheijen
1997).  One of the greatest advantages of this sample is that the radial
distribution of the old stellar population is determined from
near-infrared (K$^{'}$-band) CCD photometry which is free from the
above-mentioned problems associated with the B-band.  The galaxies are
generally of quite late morphological type and the effects of a separate
bulge component with a spheroidal shape and separate M/L are minimal. 
Indeed, the M/L in the near-infrared is probably similar for the disk
and the presumably older bulge component. 

It is important to bear in mind that this is an {\it optically} selected
sample; these galaxies are not selected for HI richness or the large
extent of the neutral hydrogen disks.  In many of these objects the
neutral hydrogen does not extend far beyond the optical disk into the
region of low accelerations where the discrepancy between the visible
and classical dynamical mass is large.  Both high and low
surface-brightness galaxies are present, as well as a range of
morphological types.  Overall, this must be considered an advantage; it
cannot be argued that these galaxies were picked as cases particularly
favorable to analysis in terms of MOND.  Moreover, the range of objects
and accelerations probed illustrate some significant general
characteristics of the mass discrepancy in spiral galaxies. 

However, this complete and homogeneous sample is not completely free
from problems.  Because it is an optically selected sample, in a number
of these objects the HI mass is low and its distribution is patchy-- not
ideal for derivation of a high-quality rotation curve.  The distance to
these galaxies is rather large compared to that of the galaxies in the
highly selected sub-sample of Paper 1.  Combined with the fact that a
number of the galaxies have a small angular size, the number of
independent points on the measured rotation curve is, in several cases,
quite low (less than five).  The relatively poor linear resolution means
that real structure in the rotation curve can be artificially smoothed
(beam smearing).  These galaxies are members of a cluster, albeit a
loose cluster, and therefore quite a number of objects are interacting
with near neighbors.  This may complicate the interpretation of the
neutral hydrogen velocity field as pure circular motion about the center
of the galaxy.  Taking these factors into account, the measured rotation
curves, as tracers of the radial force law, are generally inferior to
those of Paper 1, although the photometry is superior and the relative
error in distance is smaller. 

With these caveats in mind, we have carried out the MOND analysis on the
galaxies in the UMa sample.  This adds between 20 and 30 rotation curves
(depending upon internal selection criteria) to the total number
considered in the context of this theory.  The results are generally
positive for modified dynamics; the predicted rotation curves agree with
the observed curves and the range of implied M/L values is reasonable. 
In the K$^{'}$ band, the scatter in M/L is strikingly small: the mean
M/L is about one in solar units with a dispersion of 30\%.  This is
entirely consistent with the scatter in the observed K$^{'}$-band TF
relation (Verheijen 1997); in the context of MOND the only source of
intrinsic scatter in the TF relation is that in M/L, assuming planar gas
disks in circular rotation.  Below we describe the sample, show the
predicted and observed rotation curves and discuss the implications for
the hypotheses of modified dynamics and dark matter. 

\section{The UMa Sample}

Tully et al.  (1996) identify 79 members of the Ursa Major cluster.  The
mean recession velocity is 950 km/s with a dispersion of 150 km/s. 
Following Tully \& Verheijen (1997), the distance to all galaxies in the
sample is assumed to be 15.5 Mpc, although there is certainly a
dispersion of one or two Mpc about this mean.  There are 62 galaxies
brighter than M$_B$ $\approx -16.5$ that form a complete sample, and all
of these have been imaged in the B, R, I, and K$^{'}$ bands (Tully et
al.  1996).  The HI velocity fields have been measured for most of these
galaxies, but only 50 are sufficiently inclined (generally more than 45
degrees) to be suitable for kinematic studies.  Of these, 20 are either
early types poor in HI, are strongly interacting with neighbors, or have
fewer than five independent points along the measured rotation curve. 
Of the remaining 30, seven objects show kinematic evidence of velocity
fields with significant deviations from circular motion due to
interactions or non-axisymmetric structure (bars).  This leaves a sample
of 23 objects which are free from obvious problems (Verheijen 1997). 

The sample of 30 (including the kinematically disturbed systems) is
listed in Table 1 along with the observed properties.  The objects are
listed in column 1 along with the morphological type in column 2.  Those
systems previously designated by Verheijen (1997) as having a disturbed
velocity field are indicated by an asterisk, and the `L' in parenthesis
by the morphological type indicates that the galaxy is in the low
surface-brightness category (LSB) as defined by Tully \& Verheijen
(1997).  The B-band and K$^{'}$-band luminosities are given in columns 3
and 4.  The radial extent of the optical disk in the K$^{'}$-band, or
more specifically, the radius containing 80\% of the near-infrared
emission, is given in column 5, and the radial extent of the neutral
hydrogen in column 6.  The rotation velocity at the outer most reliable
point is given in column 7 and the centripetal acceleration in units of
$10^{-8}$ cm/s$^2$ in column 8.  We notice that in several cases, the
neutral hydrogen does not extend beyond the visible disk (e.g.  NGC
3877, NGC 3949 ); thus, these are not extended rotation curves as
considered in Papers 1 and 2.  Moreover, for several galaxies the
centripetal acceleration at the outer most point is higher than the
previously determined value (Paper 1) of the MOND acceleration parameter
($a_o=1.2\times 10^{-8}$ cm/s$^2$); therefore, in the context of MOND
one would not expect a large discrepancy in these systems. 

In several of the earlier type galaxies the mean radial distribution of
surface-brightness does show evidence for a central bulge component or
at least a central light concentration.  However, we assume that, in the
near-infrared, the difference in M/L between the bulge and disk is
small, and the light distribution is not decomposed into separate bulge
and disk components as in Papers 1 and 2 (this is justified by the
results obtained below). 

\section{Procedure and results}

The procedure for calculating and fitting MOND rotation curves to the
observed curves is exactly as described in Papers 1 and 2.  It is
assumed that the azimuthally averaged radial profile of K$^{'}$-band
emission precisely traces the stellar mass distribution in an
infinitesimally thin, axisymmetric disk; that is to say, we assume that
there is no variation of the stellar M/L within a given galaxy, and the
entire galactic mass, including any possible bulge component, is
distributed in a thin disk.  The gas mass distribution is assumed to lie
in a thin disk and to be traced by the mean radial distribution of
neutral hydrogen.  The neutral hydrogen surface density is increased by
a factor of 1.3 to account for the contribution of helium.  Thus given
the distribution of observable matter, the Newtonian gravitational
acceleration ${\bf g_n}$ is calculated in the usual way (see Begeman
1987). 

The modified dynamics posits that the true gravitational acceleration
${\bf g}$ is related to the Newtonian acceleration as $$\mu(g/a_o){\bf
g} = {\bf g_n} \eqno(1) $$ where $a_o$ is the MOND acceleration
parameter and $$\mu(x) = x(1+x^2)^{-{1\over 2}} \eqno(2) $$ which is the
commonly assumed form having the required asymptotic behavior; i.e.,
gravity is Newtonian at high accelerations but is of the MOND form $g =
\sqrt{g_na_o}$ at low accelerations (Milgrom 1983a). 

The predicted rotation curve is determined by setting the true
gravitational acceleration equal to the centripetal acceleration as
usual; i.e., $$V^2 = r\,g(r) \eqno(3)$$ This is then fit in a
least-square program to the observed rotation curve by adjusting the
mass of the luminous stellar disk, the one free parameter of the fitting
algorithm.  The acceleration parameter is not allowed to be free but is
taken to be the value determined from the fits to the high quality
rotation curves of Paper 1; i.e., $a_o = 1.2\times 10^{-8}$ cm/s$^2$. 

The distance is fixed at the value of 15.5 Mpc.  Tully and Verheijen
(1997) state that this distance is compatible with $H_o = 85$ km/s-Mpc
given the complex velocity field in the local super-cluster.  Taking a
mean radial velocity for the cluster of 1000 km/s, this distance would
also be compatible with a smooth Hubble flow and $H_o = 65$ km/s-Mpc. 
The value of $a_o$ derived in Paper 1 was based upon galaxy distances
estimated by taking $H_o = 75$ km/s-Mpc; thus, it is not immediately
evident that this same $a_o$ is consistent with the adopted distance to
the UMa cluster.  Therefore the MOND rotation curve fitting was repeated
taking distance as a free parameter in addition to M/L.  Eliminating 10
extreme values (resulting from the attempt of the least square program
to accommodate all points on possibly beam-smeared curves) the mean
distance is found to be 15.2 $\pm 3$ Mpc.  Thus the assumed distance
would seem to be consistent with the previously determined value of
$a_o$ (and with $H_o = 75$ km/s-Mpc). 

It should be noted that, in the context of MOND, the internal dynamics
of a system is affected by the external acceleration field; i.e., when
the external acceleration becomes comparable to or larger than $a_o$,
the internal dynamics is Newtonian, even though the internal
accelerations may be smaller than $a_o$ (Milgrom 1983a, b).  Because the
galaxies in this sample are members of a cluster, the magnitude of this
external field effect should be considered.  Given that the velocity
dispersion of galaxies in the UMa cluster is 150 km/s and that the
characteristic size of the cluster is 1 Mpc, we find that the cluster
acceleration on galaxies is typically on the order of $6\times 10^{-3}$
$a_o$; therefore, the individual galaxies may be considered as isolated
systems because the centripetal accelerations at the outer-most measured
point of the rotation curve are typically more than 100 times greater
than the cluster acceleration field (Table 1).  Note that the low
internal acceleration within the cluster would, in terms of MOND, imply
a large conventional virial discrepancy.  However, the dynamical
time-scale is longer than the Hubble time; the cluster is not
virialized. 

The results are given in Fig.\ 1 and Table 2.  As in Paper 2, the
observed curve is shown by the points with error bars; the predicted
MOND curve is given by the solid line; the Newtonian rotation curve of
the stellar disk corresponding to the MOND fit is given by the dotted
line; and the Newtonian curve of the gaseous disk is shown by the dashed
curve.  The observed rotation curves are determined from the velocity
field by the method of fitting tilted rings (see Begeman 1987) and
interactively adjusted for beam-smearing effects given the actual
position-velocity diagrams in the inner regions.  The error bars are
estimated by eye also from the position-velocity diagrams and probably
give a more realistic indication of the possible range of rotation
velocity than do the formal errors returned by ring-fitting program.  As
in Table 1, the asterisk indicates those systems with disturbed velocity
fields. 

Table 2 lists the fitted disk masses and errors for the sample galaxies. 
Also shown are the implied stellar mass-to-light ratios in the blue and
near infrared (columns 4 and 5), as well as the total mass-to-light
ratios in the near-infrared in column 6 (the total mass is the sum of
the fitted stellar disk mass plus the observed gas mass).  Because
galaxy evolution models predict a relation between the mass-to-light
ratio in the various bands and the color of the stellar population, the
extinction corrected B-V color is given in column 7 for those galaxies
for which it has been determined (as tabulated in the Third Reference
Catalog of de Vaucouleurs et al.  1991). 

In general, the MOND rotation curves agree well with the observed
curves, but in some cases the agreement is less than perfect.  As it
turns out, in most of those cases of serious disagreement there are
problems with the observed rotation curve or with its interpretation as
a tracer of the radial force distribution.  Several of the individual
cases are discussed below. 

NGC 3949: Verheijen (1997) notes that this rotation curve has a
considerable side-to-side asymmetry: it rises more steeply on the
receding side than on the approaching side, and there is a faint
companion 1.5 arc min to the north which may be interacting with this
galaxy. 

NGC 3992: This barred system is the most luminous galaxy in the sample. 
Although the MOND rotation curve agrees well with the observed rotation
curve, the required mass-to-light ratio of the stellar disk (2.2 in the
near-infrared) is unusually large for this sample.  This is also true
for the various dark halo fits to the rotation curve (Verheijen 1997). 

NGC 4389: This system is strongly barred, and the neutral hydrogen is
not extended but contiguous with the optical image of the galaxy. 
Verheijen (1997) points out that the velocity field cannot be
interpreted in terms of circular motion and that the overall kinematics
is dominated by the bar. 

UGC 6446: This low surface-brightness, gas-rich galaxy has an asymmetric
rotation curve in the inner regions; on the receding side it rises more
steeply than on the approaching side.  The MOND fit is much improved if
the distance to this galaxy is only 8 or 9 Mpc instead of the adopted
15.5 Mpc.  Such a possibility is consistent with the fact that this
galaxy has the lowest systemic velocity in the sample: 730 km/s which is
1.5 sigma below the mean of 950 km/s. 

UGC 6818: This is a dwarf galaxy which is probably interacting with a
faint companion on its western edge (Verheijen 1997)

UGC 6930: This is the one galaxy in the sample with an inclination less
than $45^o$ (i = $32^o$).  It was included here because of its hydrogen
richness ($M_{HI}/L_B = 0.5$) and its extremely regular velocity field. 

UGC 6973: Verheijen (1997) notes that this galaxy is interacting with
UGC 6962 to the northwest and that the HI disk is warped.  Moreover,
there is considerable evidence for vigorous star formation in the inner
region which is bright red and dusty.  In the central regions this is
the reddest galaxy in the sample; in terms of central surface brightness
${\mu_o}^B - {\mu_o}^{K'} = 6.47$ (Tully et al.  1996).  This suggests
that the K$^{'}$ band may be contaminated by dust emission and not be a
true tracer of the distribution of the old stellar population.  The
resulting calculated Newtonian rotation curve would be unrealistically
declining as a result. 

In a number of cases the calculated rotation curve overshoots the
observed rotation curve in the inner regions (e.g.  NGC 3877, NGC 4085,
NGC 4217).  These are highly inclined galaxies, and the true rotation
curve could be steeper than that measured because of missing gas in the
inner regions (HI holes).  But there is additional effect which may also
be important.  In calculating the Newtonian force from the observed
light distribution it was assumed that the stellar mass, including any
possible bulge component, was distributed entirely in a thin disk.  If
the bulge component is spheroidal this would then lead to an
over-estimate of the Newtonian force in the region of the bulge, given
that the M/L of the bulge is taken to be the same as that of the disk. 
Therefore, in several cases it might be of interest to decompose the
light distribution into a bulge and disk component.  This would not add
another fitting parameter because, in the near-infrared, M/L of the
bulge and disk is probably quite similar, but it might well improve the
fits in the inner regions. 

In addition to the usual high surface-brightness galaxies with extended
rotation curves, this sample includes low and high surface-brightness
galaxies with measured rotation curves which do not extend far beyond
the optical disk.  Because of this certain general trends are
noticeable.  For those galaxies in which the acceleration at the last
measured points of the rotation curve is comparable to $a_o$ (e.g., NGC
3877, NGC 3953, NGC 4085), the discrepancy between the Newtonian
rotation curve and the observed rotation curve is small-- exactly as
MOND predicts.  For galaxies in which the gravitational acceleration at
the last measured points is small in terms of $a_o$ (e.g., NGC 3769, NGC
4013, NGC 4157, all low surface-brightness galaxies), the observed
discrepancy is large-- again exactly as MOND predicts.  It is
interesting that in the high surface-brightness galaxies with extended
rotation curves and a large discrepancy in the outer regions (NGC 3769,
NGC 3992), Newton adequately explains the rotation curve in the inner
regions and MOND takes care of the rest. 

\section{Global mass-to-light ratios in the near-infrared and blue}

Fig.\ 2 shows the distribution of mass-to-light ratios of the stellar
disk in the near-infrared (Table 2, column 5) for the total sample of 30
galaxies.  The shaded histogram applies only to those 23 objects without
an obvious disturbance in the velocity field (the objects not designated
by an asterisk).  It is evident that the M/L$_{K^{'}}$ is sharply peaked
between 0.8 and 1.0; in the purified sample of 23 the mean M/L$_K^{'}$
is 0.97 $\pm$ 0.36.  Excluding the most extreme case, the luminous
spiral NGC 3992, this becomes 0.92 $\pm$ 0.25; i.e., there is less than
30\% scatter in the fitted near-infrared M/L. 

In Paper 2 it was pointed out that there is a rough correlation between
the MOND stellar mass-to-light ratio in the B-band and the asymptotic
rotation velocity: the more massive galaxies have a larger $M_d/L_B$. 
In Fig.\ 3, which is a plot of $M_d/L_B$ vs.  rotation velocity at the
outermost measured point, this correlation is also evident for the
galaxies in the present sample.  The filled points are the 30 galaxies
in the UMa sample, and the open points are the published sample of 33
field galaxies considered in Papers 1 and 2.  The values for the UMa
galaxies range between 0.2 and 5 and there is a clear trend with
rotation velocity of the form $M_d/L_B \propto V^3$.  The overlapping of
the two sets of points indicates the consistency of the M/L$_B$ values
here with the results for the earlier sample although the range in
M/L$_B$ and the scatter about the correlation with $V_r$ is smaller for
the UMa sample. 

More revealing is the relation between the implied stellar $M_d/L_B$ and
the reddening corrected B-V color index for those 17 sample galaxies
(Table 2) with cataloged color index data (de Vaucouleurs et al.  1991). 
This is compared again with the previous sample in Fig.\ 4 (there are
similar trends with B-R and B-K$^{'}$).  Also shown on this plot is the
predicted relation between mass-to-blue light ratio and B-V color from
the old galaxy evolution models of Larsen and Tinsley (1978); these
predicted properties are those of a population of stars evolved for
$10^{10}$ years with various prescriptions for a monotonically
decreasing star formation rate.  Although modern galaxy evolution
programs are more sophisticated (e.g.  Worthey 1994) and include
additional parameters such as the effect of varying metallicity, the
overall trends of M/L with color are the same.  Here, as previously, the
clear trend of increasing $M_d/L_B$ with redness matches that of the
model.  The evident agreement, not only in form but also in amplitude,
is somewhat fortuitous but it does indicate that the implied MOND
$M_d/L_B$ values are astrophysically plausible and consistent with
stellar population synthesis models. 

It should be recalled that the mass of the stellar disk is the only free
parameter in these rotation curve fits.  As emphasized in Paper 2 all
inadequacies in the data (dispersion in distance, deviations from
circular motion, uncertainties introduced by warping or errors in
inclination, the unknown contribution of molecular gas to the mass
distribution, etc.) are absorbed by this one parameter.  In view of
this, it is quite striking that the implied mass-to-light ratios have
characteristic value of about one in the near-infrared (with small
scatter) and in the B-band a dependence with color which is
understandable in terms of population synthesis models.  This lends
support not only to MOND but also to the assumptions underlying this
procedure (such as the constancy of M/L in a single galaxy and the
absence of a significant contribution to the surface density by
molecular gas which is distributed differently from the stars).  In
fact, the implied constancy of $M_d/L_{K^{'}}$ elevates the MOND fits to
these rotation curves to the level of actual predictions.  If one
assumes a mass-to-light ratio of about 0.9 in the near-infrared, then,
with modified dynamics, the rotation curve derived from the mean radial
distribution of near-infrared emission, when combined with the directly
observed gas distribution would be consistent, in most cases here, with
the observed rotation curve to considerable precision.  The same would
be true if one used the measured B-V color of a galaxy to estimate
$M_d/L_B$ from the Larson and Tinsley population synthesis models.  This
removes all free parameters and the rotation curve can be directly
calculated from the observed distribution of matter. 

\section{The Tully-Fisher relation}

Modified dynamics implies a mass-rotation velocity relationship for 
galaxies which is exact (Milgrom 1983a); it is the MOND equivalent of
Kepler's third law in the limit of low acceleration.
>From eqs.\ 2 and 3 it is evident that, as the radius becomes large,
$$V^4 = GM_ta_o \eqno(4) $$
where $M_t$ is the total mass of the galaxy-- the mass in gas
in addition to the stellar disk (i.e. $M_t = M_d+M_g$).  
This then leads to an observed luminosity-rotation velocity relation
(the Tully-Fisher relation) of the form
$$L = {(Ga_o<M_t/L>)}^{-1} V^4 \eqno(5) $$
Where $<M_t/L>$ is the mean total mass-to-light ratio in the particular band.
Expressed as a log(L) vs. log(V) relation, the Tully-Fisher relation
in a given color band predicted by MOND becomes
$$log(L) = a\,\,log(V) + b \eqno(6) $$
where $a = 4$ (assuming no systematic variation of 
M$_t$/L with V) and $b = -8.2 - log(<M_t/L>)$;
here $<M_t/L>$ is expressed in solar units, and we take
$a_o = 1.2 \times 10^{-8} {\rm cm/s^2}$ as found
in Paper 1.  The total mass-to-near
infrared luminosity ratio is given in column 6 of Table 2.  Here it is
found that, for the total sample $<M_t/L_{K^{'}}> = 1.2 \pm 0.56$ which means
that $b = 8.28$.  Note that MOND not
only predicts the slope of the logarithmic TF relation but also the
intercept.

The observed TF relation in the K$^{'}$ band is shown in Fig.\ 5 where
the luminosity and rotation velocity are taken directly from Table 1.  A
least-square fit, also shown on the plot, gives $a = 3.91 \pm .18$ and
$b = -8.17 \pm .39$ in agreement with the MOND prediction.  This is also
consistent with a previous determination of the H-band (1.6 $\mu$m) TF
relation for UMa galaxies by Peletier \& Willner (1993) who found a
slope of 4.08 $\pm$ 0.24.  The intrinsic scatter in the TF relation can
only arise from the intrinsic scatter in $M_t/L_k$.  Leaving out the
most extreme point, UGC 6446 (which, as noted above, may be a foreground
object), this is found to be 37\% in agreement with the scatter in the
observed K$^{'}$-band TF relation. 

The rotation velocity in Fig.\ 5 is that at the outer most reliable
point of the rotation curve and may not correspond to the asymptotic
circular velocity-- particularly in those galaxies where outermost
centripetal accelerations are greater than $a_o$ and in the low
surface-brightness galaxies with rotation curves still rising at the
last point.  Verheijen (1997) has noted that by selecting only those
galaxies in the Ursa Major sample in which the rotation curve has
reached an asymptotically constant value, and by taking this asymptotic
value for the velocity coordinate in the TF relation (as opposed to a
global profile width or the peak rotation velocity), the scatter in the
TF relation is greatly reduced and the slope is four (see also Broeils
1992, Rhee 1996, and McGaugh \& de Blok 1998a).  This is precisely the
expectation deduced from MOND where it is the constant rotation velocity
at large distance from the visible galaxy which correlates exactly with
mass (eq.\ 4).  Verheijen further notes that in this highly selected
sub-sample of fifteen galaxies the scatter in the observed near-infrared
TF relation is consistent with the observational scatter; that is to
say, no intrinsic scatter in the TF relation can be detected.  This is
also consistent with MOND in which the mass-velocity relation is exact:
the only source of intrinsic scatter is that in M$_t$/L which, in the
near-infrared, may be quite small. 

Twelve of the galaxies in this sample are in the low surface-brightness
category as defined by Tully and Verheijen (1997); these are mostly the
lower luminosity objects in Fig.\ 5 although there is some overlap in
luminosity with objects of high surface-brightness.  It is noteworthy
that these LSB galaxies lie on the same TF relation defined by the high
surface-brightness galaxies.  This fact, which has been noticed by Zwaan
et al.  (1995), is an inevitable aspect of MOND where the TF-relation is
absolute (McGaugh \& de Blok 1998b).  Dark halo models, on the other
hand, require some contrived coupling of parameters to reproduce this
observation (McGaugh \& de Blok 1998a). 

\section{Conclusions}

This sample of galaxies in the Ursa Major cluster combined with those of
Paper 2 and the low surface-brightness galaxies analyzed by McGaugh and
de Blok (1998b), provide a total sample of about 80 spiral galaxies with
rotation curves measured in neutral hydrogen for which the predicted
MOND rotation curves have been calculated.  Because of the systematic
effects the usual statistical tests of goodness-of-fit are not entirely
appropriate in accessing these rotation curve fits, but qualitatively
one would say that in only five or six cases out of this 80 is the
rotation curve predicted by MOND from the observed distribution of light
and gas noticeably different from the observed rotation curve.  In the
present sample, NGC 4389 and UGC 6973 should be included in this
category.  For these objects there is usually an obvious problem with
the observed rotation curve or its use as a tracer of the radial force
distribution.  In others one may question the measured surface
brightness distribution as a precise tracer of the stellar mass
distribution (UGC 6973 for example).  For about 30 galaxies the MOND
rotation curve reproduces the general shape of the observed curve and
accounts for the magnitude of the conventional mass discrepancy with a
reasonable mass-to-light ratio (NGC 3877, 4013, 4138 would fall in this
category).  In fully half of the objects considered, the MOND rotation
curve reproduces the observed curve with considerable precision (as for
NGC 4157, UGC 6399, 6667, 6917 in the present sample). 

Unlike the previous samples, the galaxies considered here are optically
selected and form a complete and homogeneous sample.  Because selection
does not depend primarily upon neutral hydrogen richness and extent,
these galaxies have a wider range of properties than those in Papers 1
and 2.  For example, in a number of these objects the observed rotation
curve does not extend far beyond the optical disk into the regime where
the discrepancy is large.  This is a positive aspect of this sample
because certain general predictions of MOND are seen to be verified: the
discrepancy between the Newtonian and observed rotation curve is seen to
be small in those objects with high centripetal accelerations at the
last measured point of the rotation curve-- as MOND predicts.  In
regions where the internal acceleration is low-- in the extended
rotation curves of high surface-brightness galaxies as in the previous
samples-- MOND predicts the observed large discrepancy.  It is
noteworthy that some of the best fits are achieved for the low
surface-brightness galaxies where the discrepancy is largest.  This is
precisely what one would expect from MOND since these objects lie
entirely in the low acceleration limit of the theory-- in the deep MOND
limit-- where there is no uncertainty arising from unknown function
$\mu(x)$ in eq.\ 2 (see also McGaugh \& de Blok 1998b). 

The range of M/L is generally reasonable and consistent, in the blue,
with population synthesis models.  When near-infrared photometry is
available, as for this sample, there is remarkably little variation in
the fitted M/L of the stellar component-- on the order of 30\%.  When
one considers that this is the only free parameter in the fits and will
reflect all uncertainties (such as the real dispersion in distance), the
results here are consistent with a effectively constant value of M/L in
the near-infrared for spiral galaxies.  Assigning the same near-infrared
M/L to all galaxies then makes the MOND rotation curves true
predictions, without free parameters. 

The predictive power of MOND, at least with respect to galaxy rotation
curves and the TF relation, is well-established.  It has been difficult
to compare this, in a fair way, with dark matter halos because, until
recently, the dark halo hypotheses have had essentially no predictive
power.  The exercise has been one of fitting dark matter halos with an
assumed density distribution to the observed rotation curve in order to
estimate halo and disk parameters.  With at least three free parameters
available (disk M/L, halo core radius, and density normalization),
essentially any observed rotation curve can be reproduced, as in the
analysis of UMa galaxies by Verheijen (1997). 

In the last several years, this has changed due to the further
development of cosmological N-body codes with large numbers of particles
and high spatial resolution.  It has become evident that in simulations
in which the initial spectrum of density fluctuations is that of Cold
Dark Matter (CDM), a characteristic form for the radial mass
distribution of virialized halos arises (Dubinsky \& Carberg 1991,
Navarro et al.  1996, Cole \& Lacey 1996).  This characteristic form is
distinguished by a singular density distribution-- a density which
continues to increase as 1/r into the origin.  This density law may be
parameterized by the Hernquist model (Hernquist 1990, Dubinsky \&
Carlberg 1991) or by an alternative model suggested by Navarro et al. 
(1996) which has essentially the same form in the regime where galaxy
rotation curves actually measured.  Although these singular halos can
produce acceptable fits to the rotation curves of high luminosity, high
surface-brightness galaxies (Sanders and Begeman 1994) they generally
fail in low surface-brightness dwarfs (Flores \& Primack 1994, Moore
1994).  Significantly, Navarro et al.  (1996) have demonstrated that,
given the mass or velocity scaling of a halo, the degree of central
concentration (or halo length scale) is determined by the cosmology. 
McGaugh \& de Blok (1998b) show that such constrained halos dramatically
fail to reproduce the rotation curves of the low surface-brightness
galaxies for plausible cosmologies; it is impossible to match the
gradual rise in the observed rotation curve in the inner region with the
asymptotic velocity at larger radii; the singular density distribution
produces a rotation curve which rises far too steeply. 

The predicted mass-velocity relation for the CDM halos (not considering
that the observed correlation is with the luminosity of the baryonic
component) is closer to $M\propto V^3$ (White 1997) while $M\propto V^4$
for MOND-- again a clearly distinct prediction.  In this sense also,
MOND is successful when one considers the correlation between the
asymptotic flat rotation velocity and the luminosity (Verheijen 1997). 
Therefore, on this very basic phenomenological level, one can only
conclude that MOND works where dark matter, at least the presently
favored form of dark matter, does not. 

It is sometimes argued that MOND is ``designed'' to fit rotation curves,
so that it is no surprise that it works so well on this scale.  It is
true that MOND, in some sense, is designed to reproduce asymptotically
flat rotation curves and a TF relation of the form $L\propto V^4$.  But
it by no means evident that the variety of detailed shapes of rotation
curves exhibited by the total sample of 80 galaxies so far considered
could be so precisely reproduced by using MOND to calculate the radial
force from the observed distribution of detectable matter.  MOND works
well throughout the entire galaxy, not just where the rotation attains
its asymptotic constant value, and it was in no sense designed to do
this.  Using MOND Milgrom {\it predicted} that the discrepancy between
the observed curve and the Newtonian rotation curve should be small in
regions of high surface-brightness, and that the discrepancy should be
large in galaxies of low surface-brightness-- even before such galaxies
were discovered.  These predictions are clearly verified in the sample
of galaxies considered here, which include both types of systems.  It is
this inherent simplicity and predictive power that gives the idea its
continued impact in spite of the absence of a solid theoretical basis
for MOND.  The detailed shape and amplitude of a rotation curve can be
calculated from the observed distribution of detectable matter, with no
free parameters by assuming a fixed value of M/L of about one in the
near-infrared.  What more could one expect from a theory of gravity?
Because of this it is justified to claim that MOND, at the present time,
is epistemologically superior to the dark matter hypothesis. 

\acknowledgements 
We are very grateful to M.\ Milgrom and S.S. McGaugh for very helpful
comments on this manuscript.

\clearpage 

\begin{table}[p]
\tighten
\begin{flushleft}
\singlespace
\caption{ The Ursa Major Galaxies \label{t1}}
\begin{tabular}{|c|c|c|c|c|c|c|c|c|}
\tableline
${\rm Galaxy }$ &${\rm Type}$ & ${\rm L_B}$ & ${\rm L_{K^{'}}}$ & 
   ${\rm R_{80}}$ 
   & ${\rm R_{HI}}$ & ${\rm M_{HI}}$ & ${\rm V_{rot}}$ & ${\rm a}$ \\
   $  $ & $ $ &   ${\rm 10^{10}\,L_\odot}$ & 
   ${\rm 10^{10}\,L_\odot}$ & ${\rm kpc}$ & ${\rm kpc}$ & 
   ${\rm 10^{10}\, M_\odot}$
   & ${\rm km/s}$ & ${\rm 10^{-8}\, cm/s^2}$  \\
 (1) & (2) & (3) & (4) & (5) & (6) & (7) & (8) & (9) \\ 
\tableline
  NGC 3726 & SBc & 2.65 & 3.56 & 7.5 & 28 & 0.48 & 162 & 0.30 \\
  NGC 3769* & SBb & 0.68 & 1.27 & 4.4 & 32 & 0.41 & 122 & 0.15 \\
  NGC 3877 & Sc & 1.94 & 4.52 & 7.0 & 10 & 0.11 & 167 & 0.90 \\
  NGC 3893* & Sc & 2.14 & 3.98 & 5.8 & 18 & 0.43 & 188 & 0.64 \\
  NGC 3917 & Scd(L) & 1.12 & 1.35 & 6.4 & 13 & 0.14 & 135 & 0.45 \\
  NGC 3949 & Sbc & 1.65 & 2.33 & 4.4 & 6 & 0.25 & 164 & 1.45 \\
  NGC 3953 & SBbc & 2.91 & 8.47 & 8.5 & 14 & 0.21 & 223 & 1.14 \\
  NGC 3972 & Sbc & 0.68 & 1.00 & 5.0 & 8 & 0.09 & 134 & 0.73 \\
  NGC 3992 & SBbc & 3.10 & 6.98 & 8.8 & 30 & 0.71 & 242 & 0.63 \\
  NGC 4010 & SBd(L) & 0.63 & 1.20 & 8.6 & 9 & 0.21 & 128 & 0.59 \\
  NGC 4013 & Sb & 1.45 & 4.96 & 5.3 & 27 & 0.22 & 177 & 0.38 \\
  NGC 4051* & SBbc & 2.58 & 3.91 & 6.4 & 11 & 0.20 & 159 & 0.74 \\
  NGC 4085 & Sc & 0.81 & 1.22 & 3.5 & 5 & 0.10 & 134 & 1.20 \\
  NGC 4088* & Sbc & 2.83 & 5.75 & 7.2 & 19 & 0.61 & 173 & 0.51 \\
  NGC 4100 & Sbc & 1.77 & 3.50 & 6.5 & 20 & 0.23 & 164 & 0.44 \\
  NGC 4138 & Sa & 0.82 & 2.88 & 3.0 & 16 & 0.11 & 147 & 0.44 \\
  NGC 4157 & Sb & 2.00 & 5.75 & 6.6 & 26 & 0.61 & 185 & 0.43 \\
  NGC 4183 & Scd(L) & 0.90 & 0.73 & 7.4 & 18 & 0.26 & 112 & 0.23 \\
  NGC 4217 & Sb & 1.90 & 5.29 & 6.9 & 14 & 0.19 & 178 & 0.73 \\
  NGC 4389* & SBbc & 0.61 & 1.22 & 3.6 & 5 & 0.04 & 110 & 0.78 \\
  UGC 6399 & Sm(L) & 0.20 & 0.21 & 4.8 & 7 & 0.05 & 88 & 0.36 \\
  UGC 6446 & Sd(L) & 0.25 & 0.14 & 4.0 & 13 & 0.23 & 82 & 0.17 \\
  UGC 6667 & Scd(L) & 0.26 & 0.28 & 7.6 & 7 & 0.06 & 86 & 0.34 \\
  UGC 6818* & Sd(L) & 0.18 & 0.12 & 3.6 & 6 & 0.08 & 73 & 0.29 \\
  UGC 6917 & SBd(L) & 0.38 & 0.42 & 5.8 & 9 & 0.15 & 110 & 0.44 \\
  UGC 6923 & Sdm(L) & 0.22 & 0.21 & 3.3 & 5 & 0.06 &  81 & 0.31 \\
  UGC 6930(i) & SBd(L) & 0.50 & 0.40 & 3.7 & 14 & 0.24 & 110 & 0.23 \\
  UGC 6973* & Sab & 0.62 & 2.85 & 1.7 & 7 & 0.13 & 173 & 1.39 \\
  UGC 6983 & SBcd(L) & 0.34 & 0.34 & 5.1 & 14 & 0.22 & 107 & 0.27 \\
  UGC 7089 & Sdm(L) & 0.44 & 0.21 & 5.1 & 8 & 0.09 & 79 & 0.25 \\
\tableline 
\end{tabular}
\end{flushleft}
\end{table}
\clearpage
\begin{table}[p]
\singlespace
\tighten
\begin{flushleft}
\caption{ The MOND mass and the implied M/L values \label{t2}}
\begin{tabular}{|c|c|c|c|c|c|c|c|c|}
\tableline
${\rm Galaxy }$ &${\rm M_d}$ & $ \pm $ & ${\rm M_d/L_B}$ & 
   ${\rm M_d/L_{K^{'}}}$ 
   & ${M_t/L_{K^{'}}} $ & $ B-V $ \\ 
   $  $ &  ${\rm 10^{10}\, M_\odot} $ & ${\rm 10^{10}\, M_\odot} $ & 
   $ $ & $ $ & $ $ & $ $ \\
 (1) & (2) & (3) & (4) & (5) & (6)  & (7) \\ 
\tableline
  NGC 3726 & 2.62 & 0.20 & 0.99 & 0.74 & 0.92 & 0.45 \\
  NGC 3769* & 0.80 & 0.05 & 1.18 & 0.63 & 1.04 &  \\
  NGC 3877 & 3.35 & 0.17 & 1.72 & 0.74 & 0.77  & 0.68 \\
  NGC 3893* & 4.20 & 0.27 & 1.96 & 1.06 & 1.19 &   \\
  NGC 3917 & 1.40 & 0.09 & 1.25 & 1.04 & 1.17  & 0.60 \\
  NGC 3949 & 1.39 & 0.16 & 0.84 & 0.60 & 0.73 & 0.39 \\
  NGC 3953 & 7.88 & 0.22 & 2.71 & 0.93 & 0.96 & 0.71 \\
  NGC 3972 & 1.00 & 0.08 & 1.47 & 1.00 & 1.12 & 0.55 \\
  NGC 3992 & 15.28 & 0.30 & 4.93 & 2.19 & 2.27 & 0.72 \\
  NGC 4010 & 0.86 & 0.07 & 1.37 & 0.72 & 0.94 & \\
  NGC 4013 & 4.55 & 0.08 & 3.13 & 0.92 & 0.97 & 0.83 \\
  NGC 4051* & 3.03 & 0.15 & 1.17 & 0.77 & 0.84 & 0.62 \\
  NGC 4085 & 1.00 & 0.13 & 1.23 & 0.82 & 0.92 & 0.47 \\
  NGC 4088* & 3.30 & 0.18 & 1.16 & 0.57 & 0.71 & 0.51\\
  NGC 4100 & 4.32 & 0.12 & 2.44 & 1.23 & 1.32 & 0.63 \\
  NGC 4138 & 2.87 & 0.25 & 3.50 & 1.00 & 1.04 & 0.81 \\
  NGC 4157 & 4.83 & 0.16 & 2.42 & 0.84 & 0.98 & 0.66 \\
  NGC 4183 & 0.59 & 0.035 & 0.65 & 0.81 & 1.27 & 0.39 \\
  NGC 4217 & 4.25 & 0.22 & 2.24 & 0.80 & 0.85 & 0.77 \\
  NGC 4389* & 0.233 & 0.081 & 0.38 & 0.19 & 0.23 & \\
  UGC 6399 & 0.207 & 0.008 & 1.04 & 0.99 & 1.33 & \\
  UGC 6446 & 0.117 & 0.018 & 0.47 & 0.87 & 2.99 & \\
  UGC 6667 & 0.249 & 0.022 & 0.96 & 0.89 & 1.17 & \\
  UGC 6818* & 0.040 & 0.013 & 0.22 & 0.33 & 1.18 & \\
  UGC 6917 & 0.541 & 0.023 & 1.42 & 1.29 & 1.75 & \\
  UGC 6923 & 0.164 & 0.030 & 0.75 & 0.78 & 1.11 & 0.42 \\ 
  UGC 6930 & 0.416 & 0.023 & 0.83 & 1.04 & 1.84 & \\
  UGC 6973* & 1.69 & 0.190 & 2.72 & 0.59 & 0.64 & \\
  UGC 6983 & 0.565 & 0.036 & 1.66 & 1.66 & 2.51 & \\
  UGC 7089 & 0.092 & 0.005 & 0.21 & 0.44 & 0.96 & \\
\tableline 
\end{tabular}
\end{flushleft}
\end{table}
\clearpage

\figcaption[ ]{MOND fits to the rotation curves of the UMa galaxies.
The radius (horizontal axes) is given in kpc in all cases and the rotation
velocity in km/s.
The points with error bars are the observations and the solid line is
the rotation curve determined from the distribution of light and neutral
hydrogen with the MOND formula.  The other curves are the Newtonian
rotation curves of the various separate components:  the 
short dashed line is the rotation curve of the gaseous disk (HI plus He);
the dotted curve is that of the luminous disk.  The free parameter of
the fitted curve is the disk mass.  \label{fig1}}

\figcaption[ ]{The distribution of the mass-to-light ratio in the
near-infrared, ${\rm M_d/L_{K^{'}}}$, for the sample of 30 UMa galaxies
where M$_d$ is the mass of the stellar disk determined from the MOND 
fit.  The shaded histogram is for the sub-sample of 23 galaxies with
undisturbed velocity fields.  \label{fig2}}

\figcaption[ ]{A log-log plot of ${\rm M_d/L_B}$ vs. the observed
asymptotic rotation
velocity for the sample galaxies (solid points) compared with the
the previous sample (open points) of field galaxies (Paper 2).  
Here ${\rm M_d}$ is the total mass of the stellar disk determined 
from the MOND fit. \label{fig3}}

\figcaption[ ]{Log of ${\rm M_d/L_B}$ of sample (solid points) galaxies vs.
the reddening corrected B-V color from the Third Reference Catalogue (de  
Vaucouleurs et al. 1991) compared with the previous sample (open points) 
of field galaxies (Paper 2).  Also shown (dashed line) is the theoretical
${\rm M/L_B}$ vs. B-V
from population synthesis models of Larson and Tinsley (1978).
\label{fig4}}

\figcaption[ ]{A log-log plot of the K$^{'}$-band luminosity
($10^{10}$ ${\rm L_\odot}$) vs. the observed
asymptotic rotation velocity (the K$^{'}$-band Tully-Fisher relation)
for the 30 sample galaxies.  The unshaded points are those galaxies
with perturbed velocity fields.  For the total sample, a least-square
fit yields $log(L) = (3.91 \pm .18) log(V) - (8.17 \pm .39)$ which is also
shown by the solid line.
\label{fig5}} 

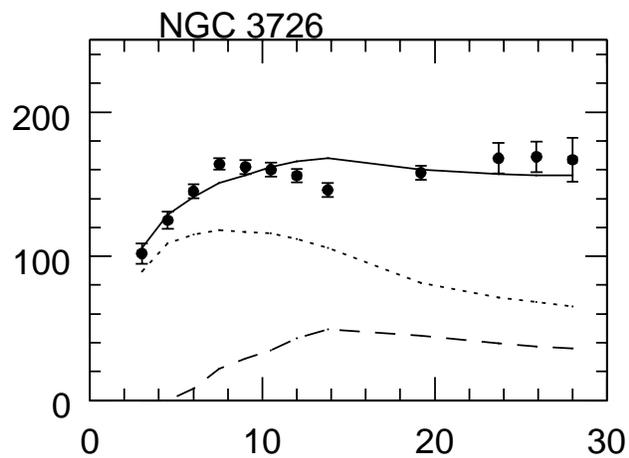
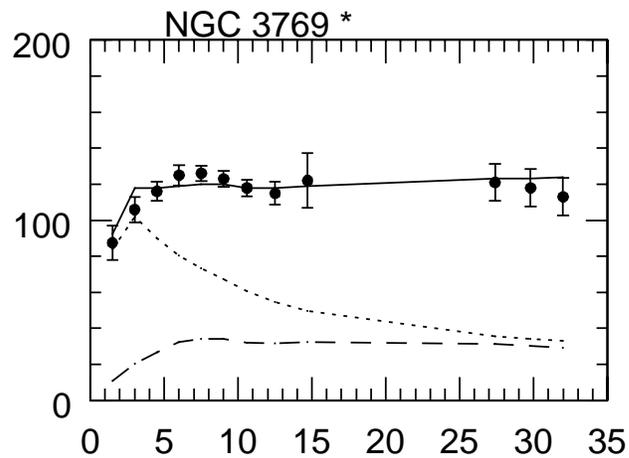
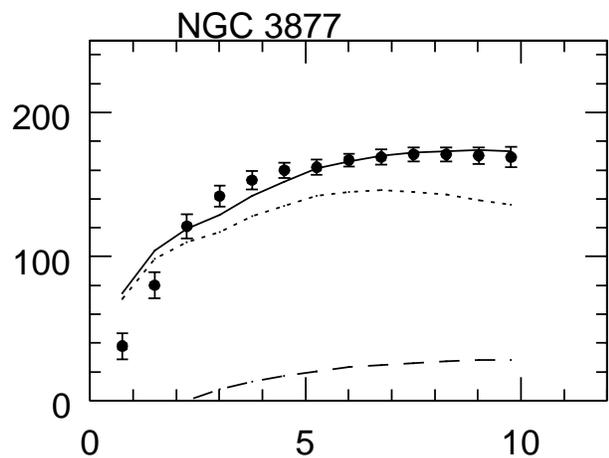
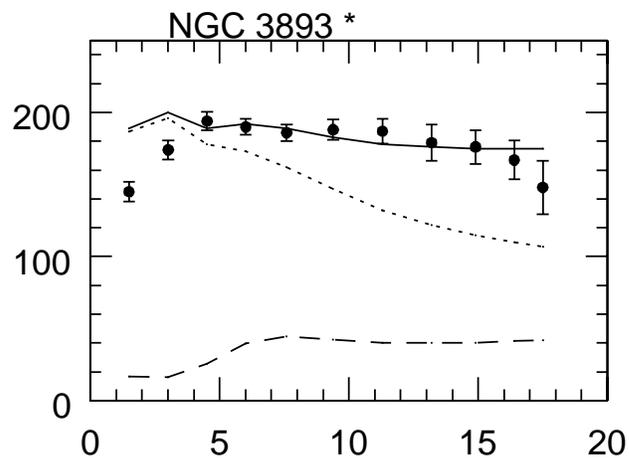
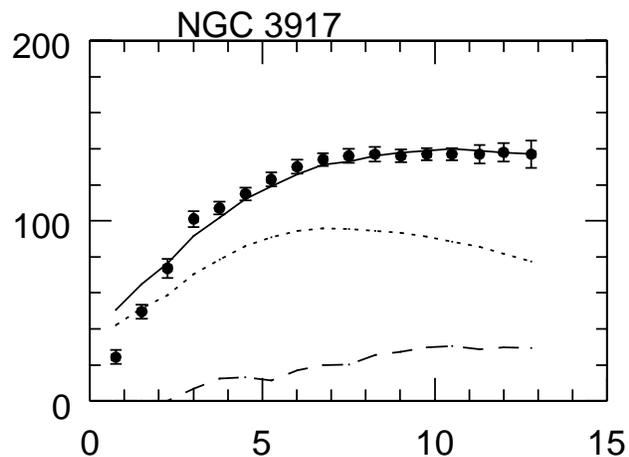
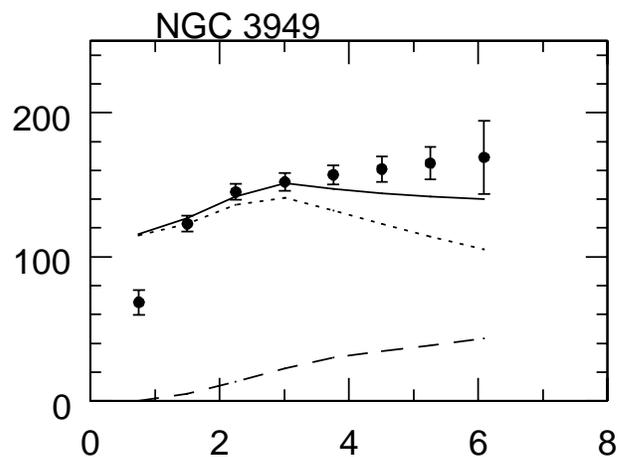
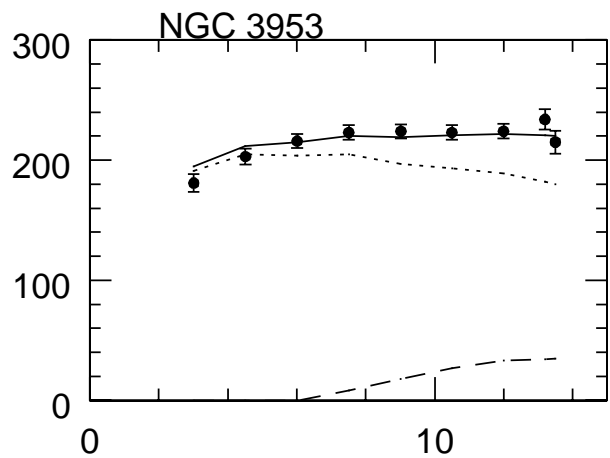
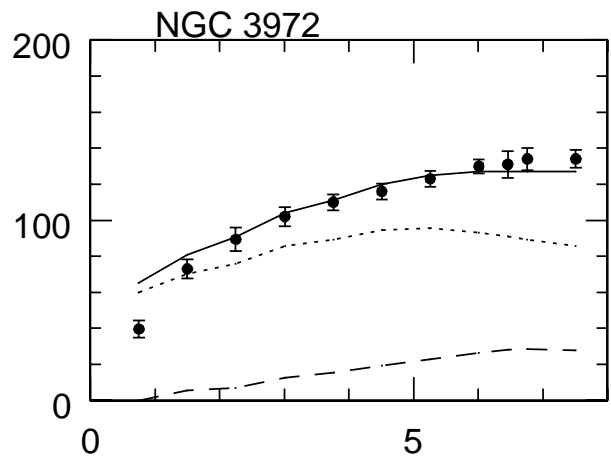
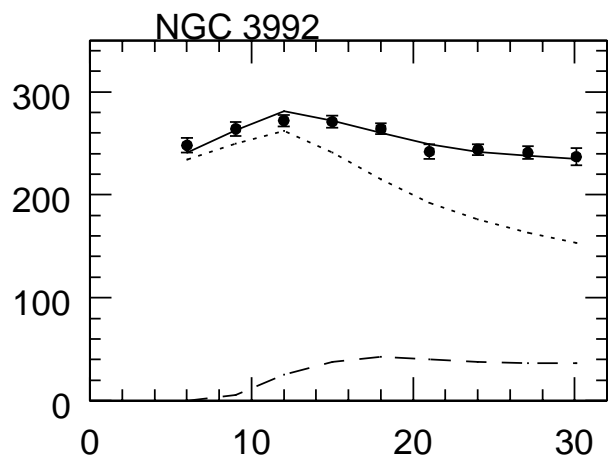
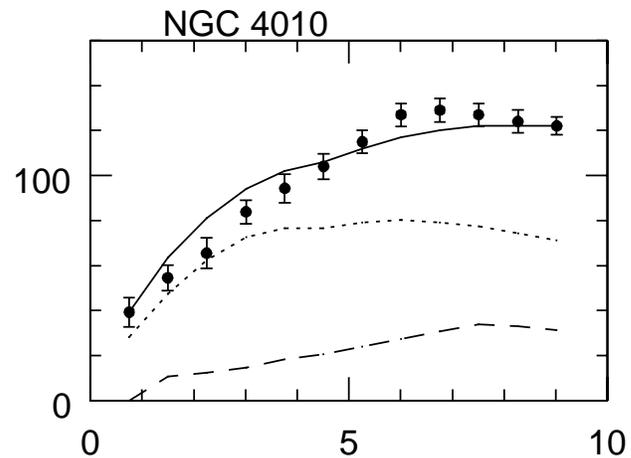
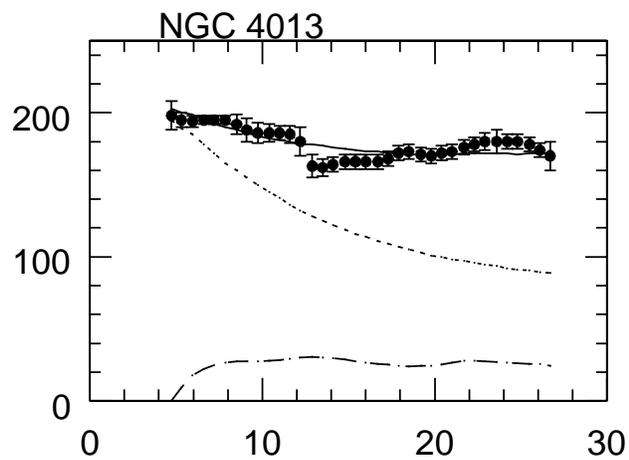
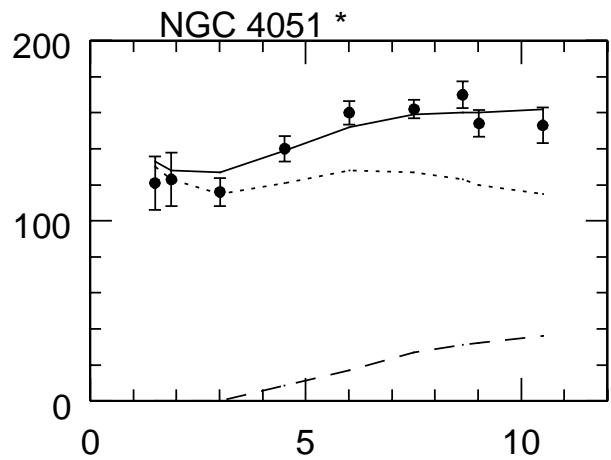
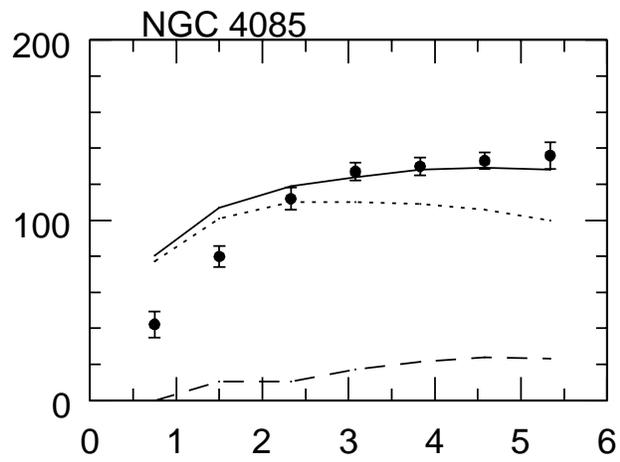
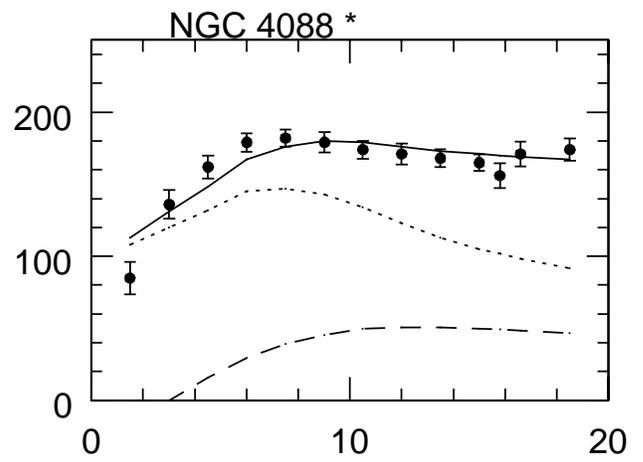
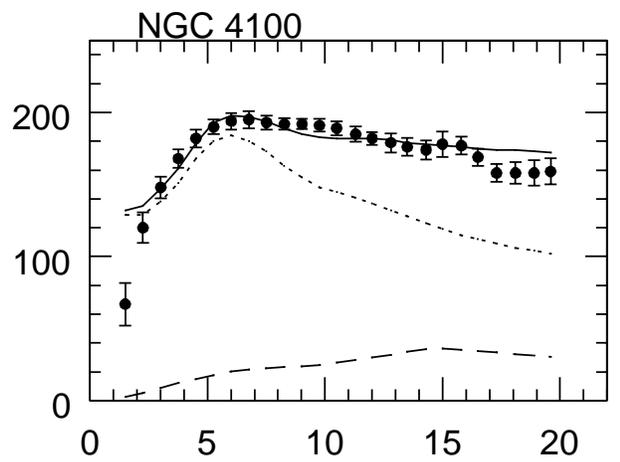
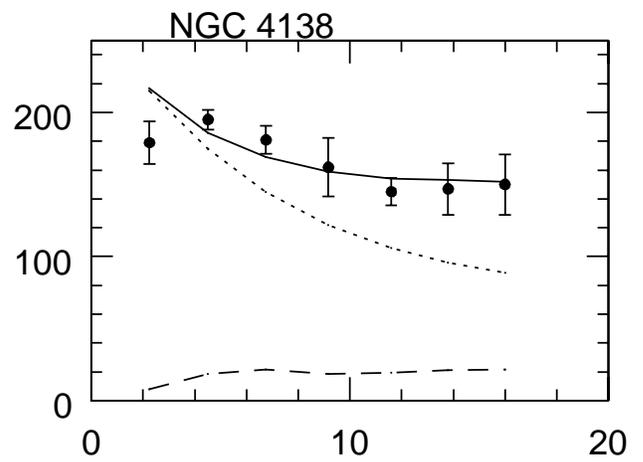
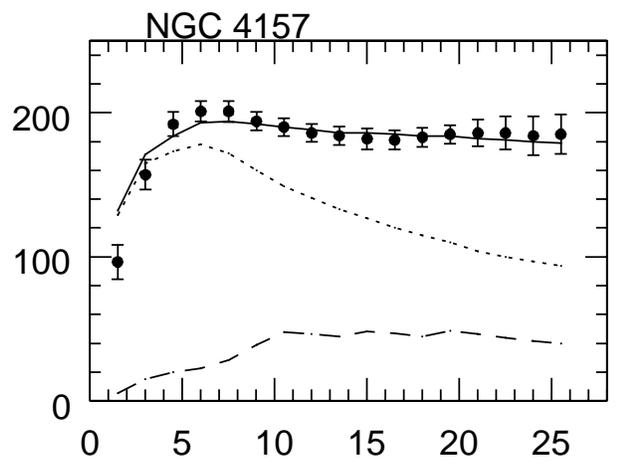
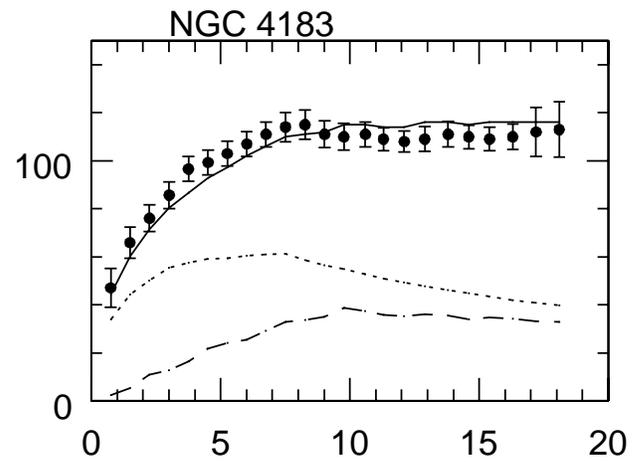
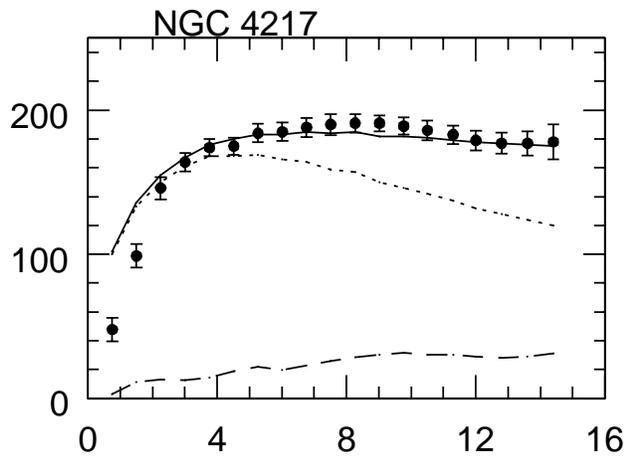
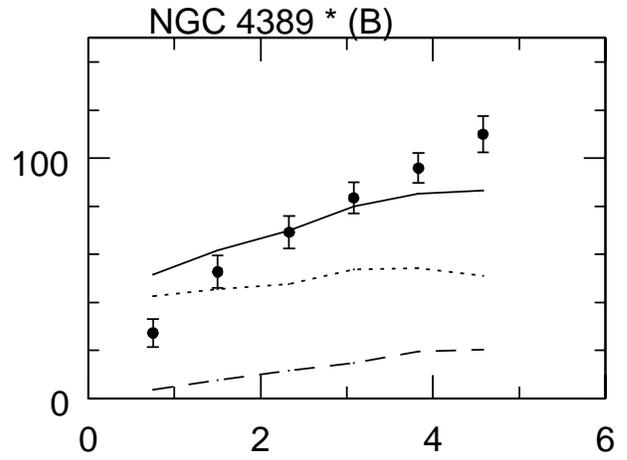
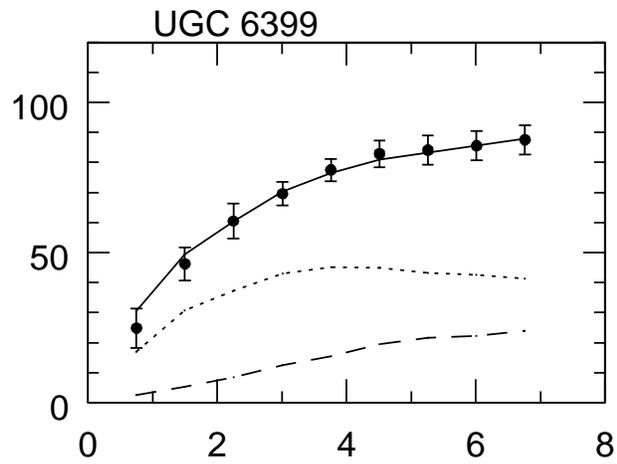
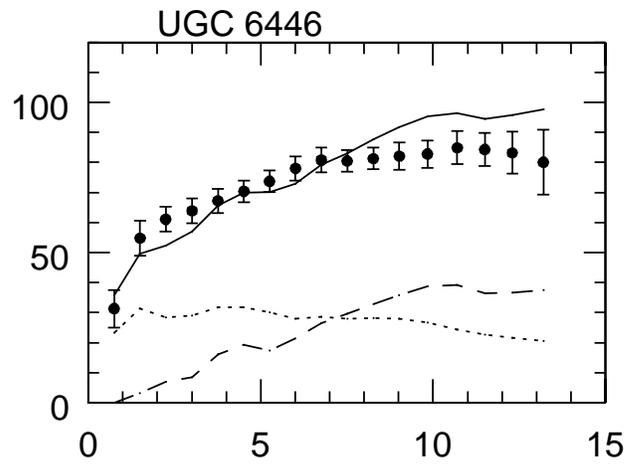
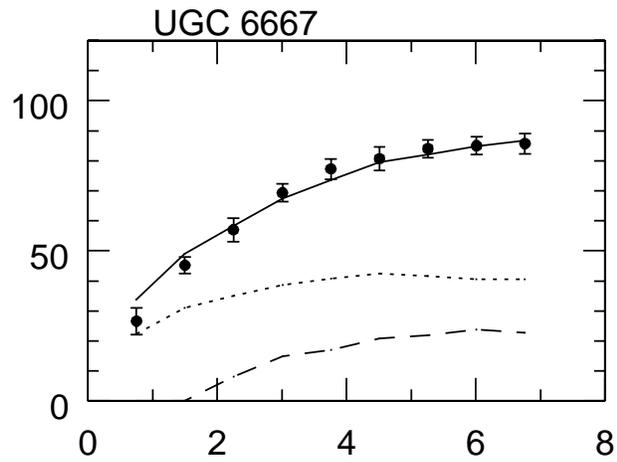
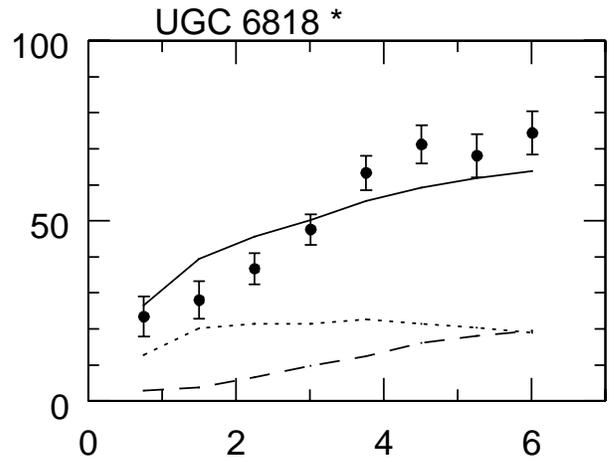
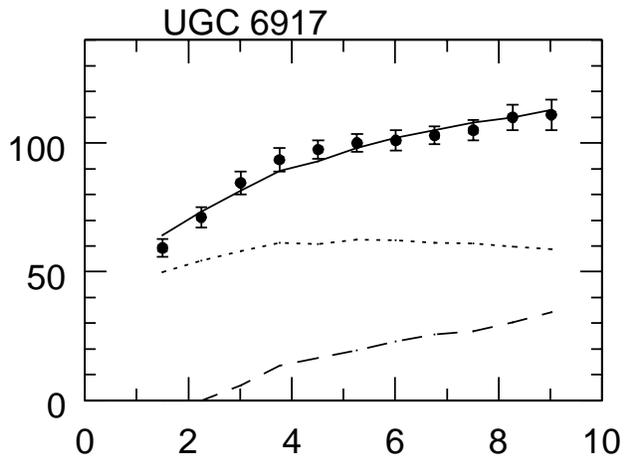
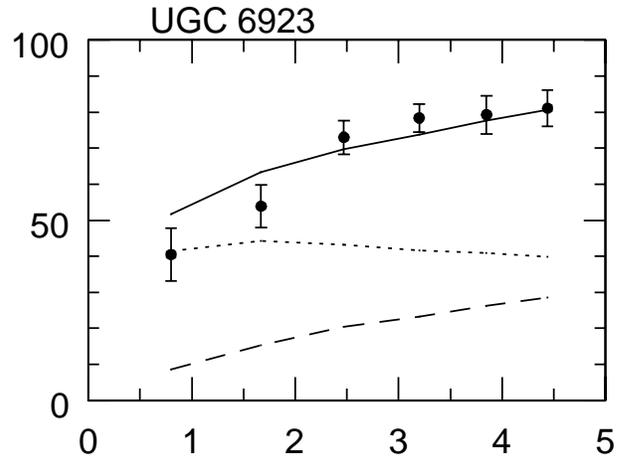
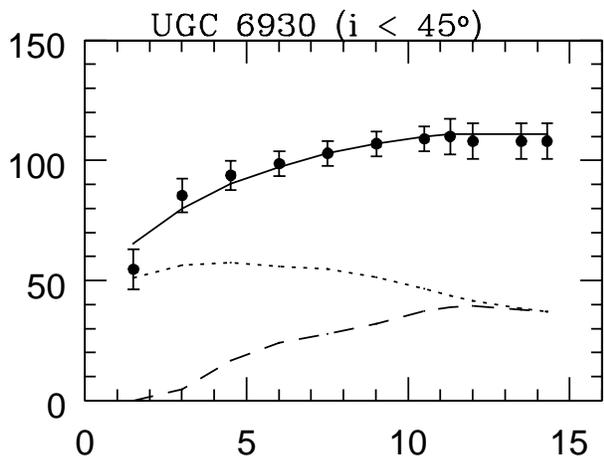
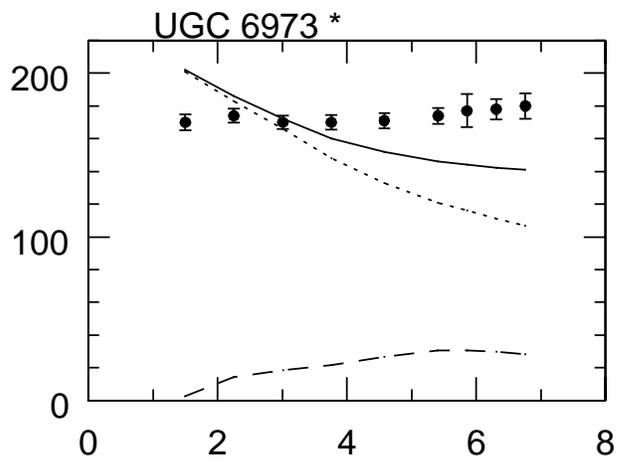
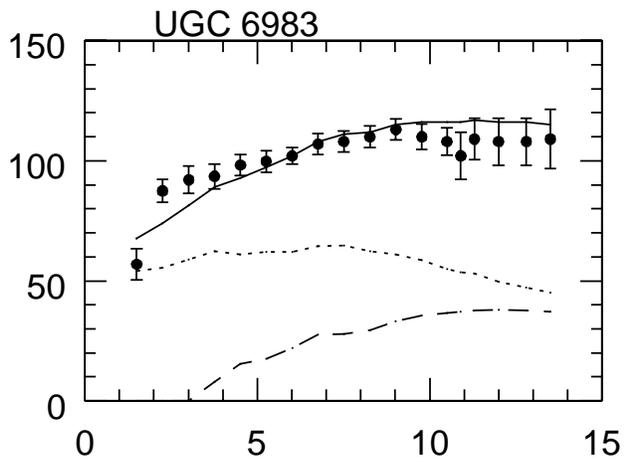
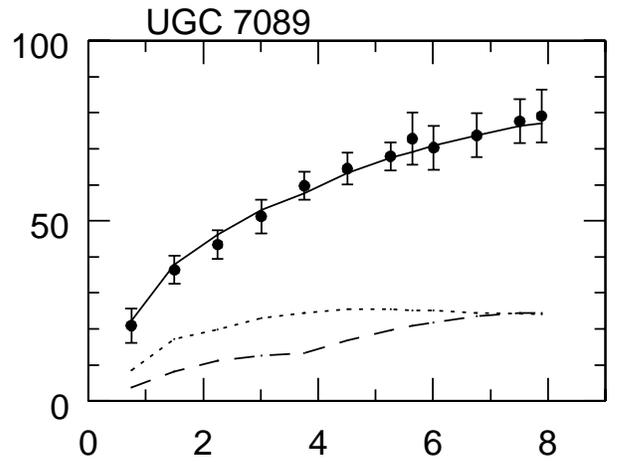
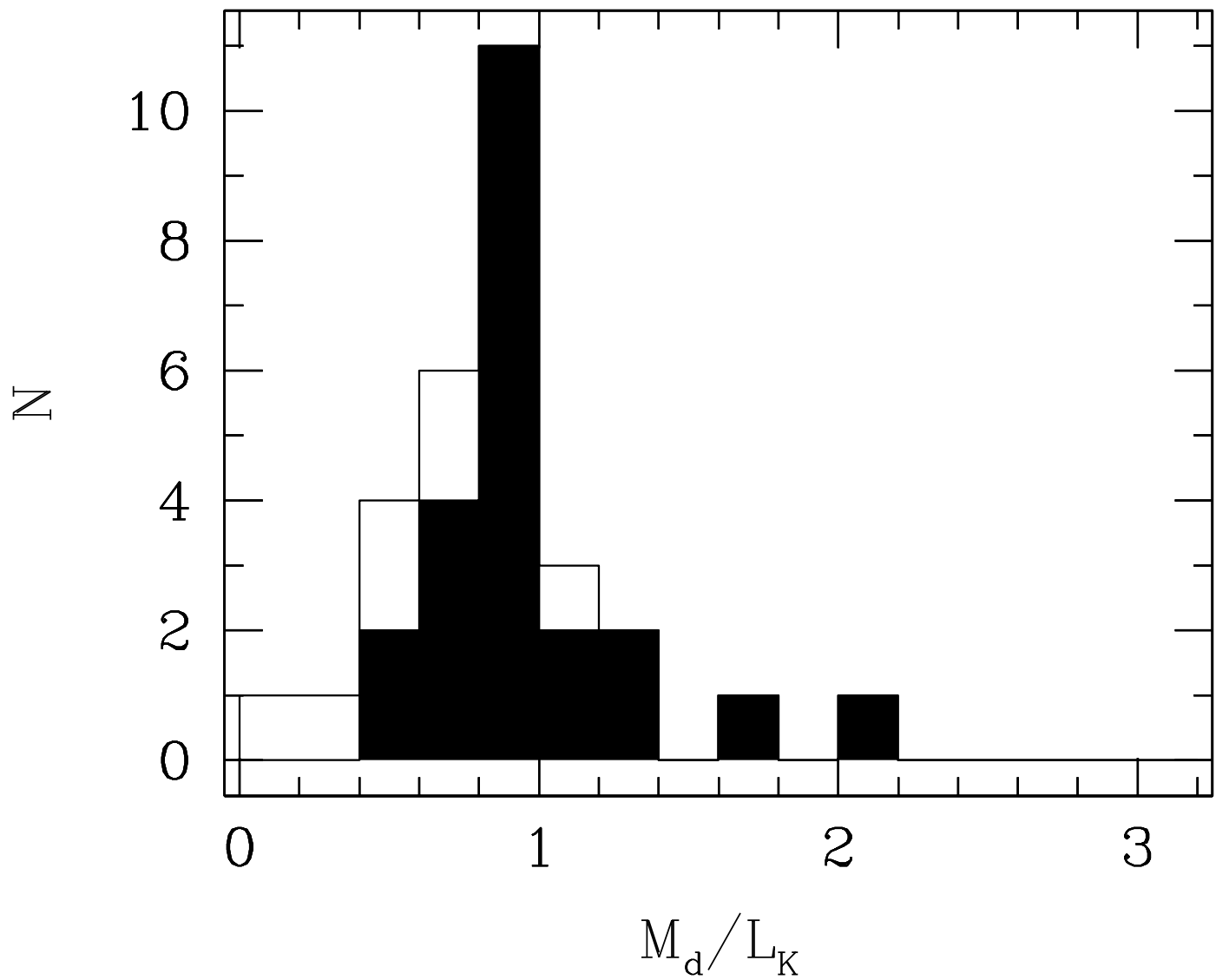
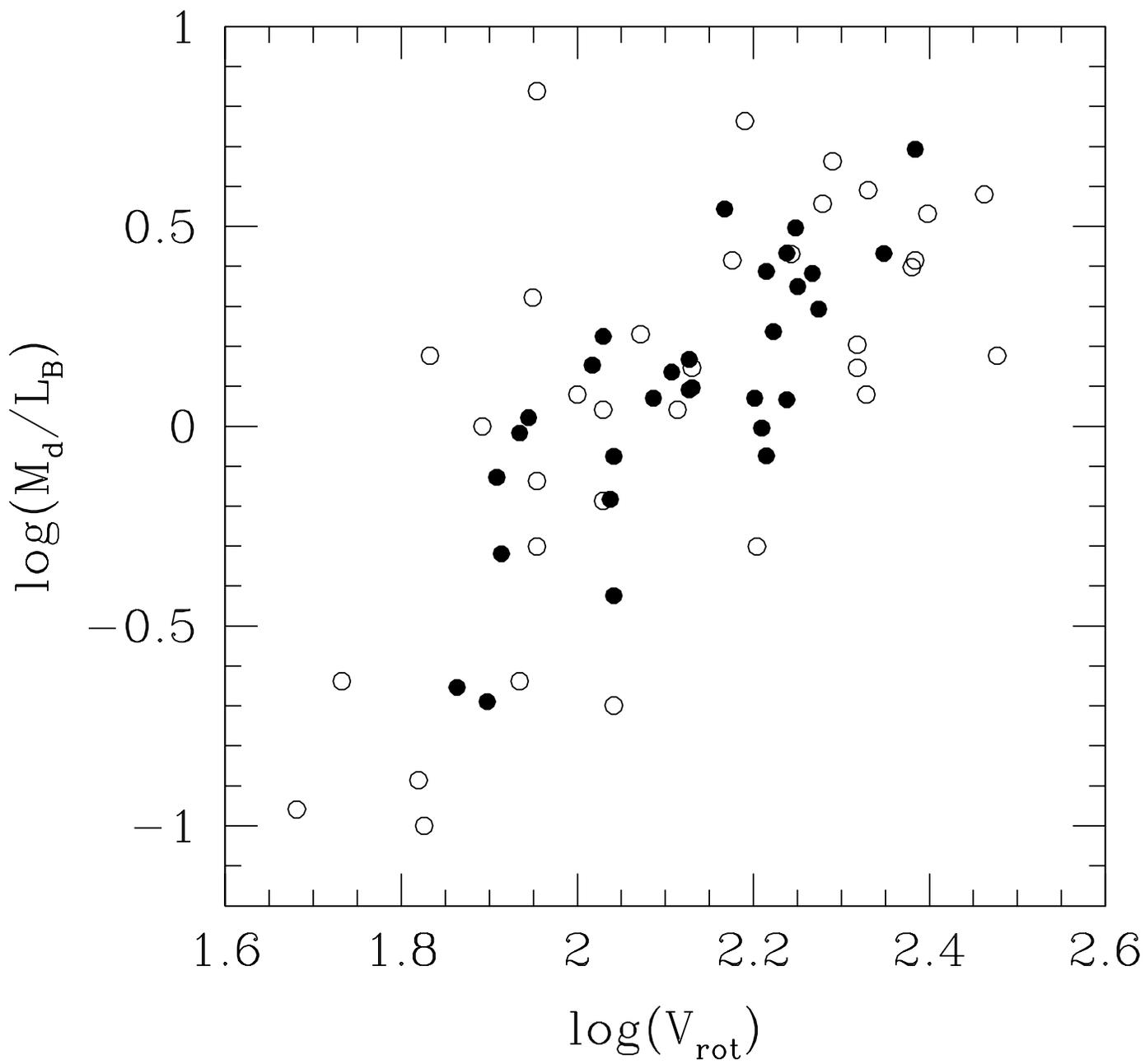
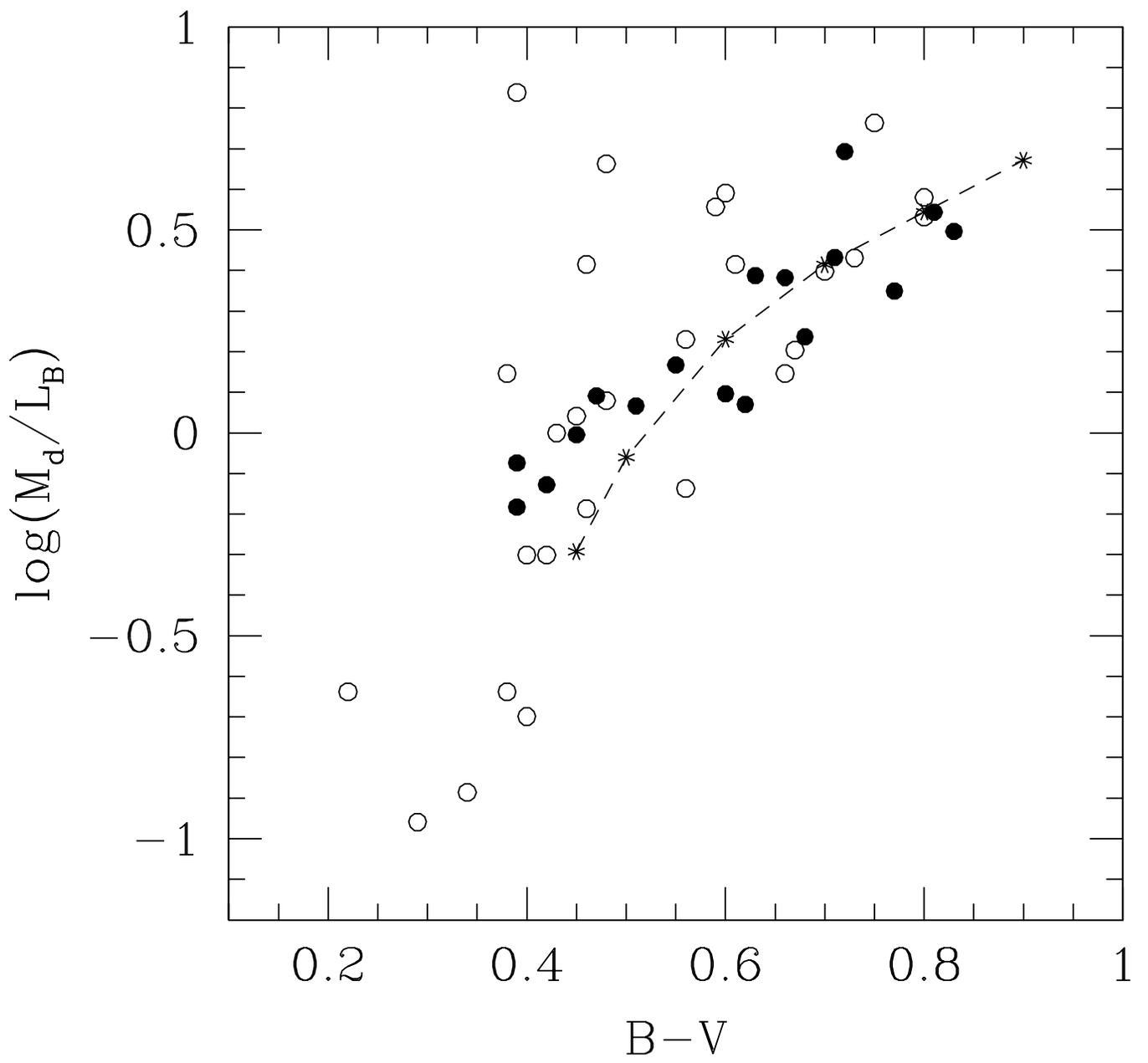
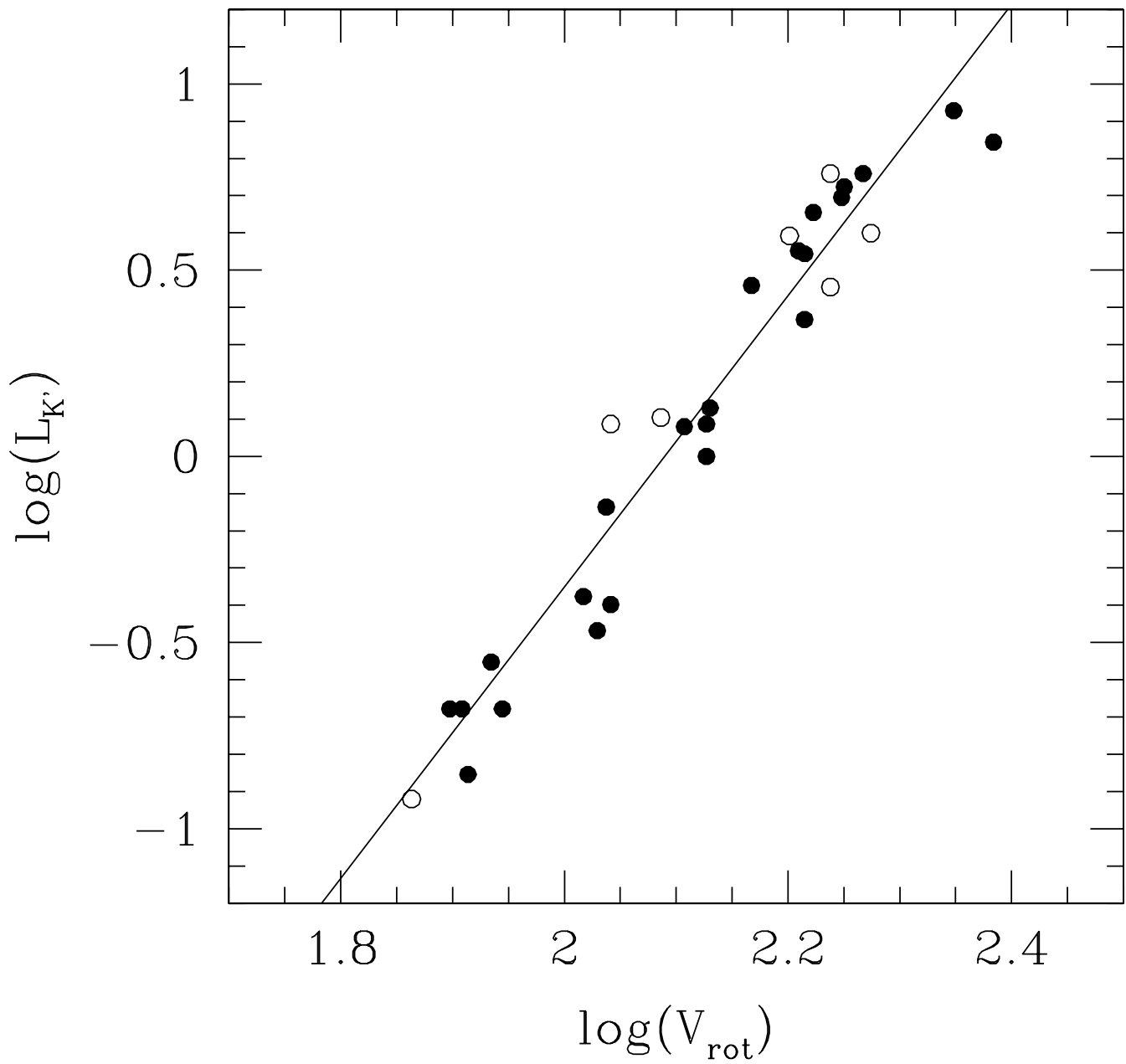
\end{document}